\documentclass[smallabstract,smallcaptions]{dccpaper}

\usepackage[T1]{fontenc}
\usepackage{graphicx}
\PassOptionsToPackage{hyphens}{url}
\usepackage{hyperref}       
\usepackage{url}
\usepackage{amsfonts}       
\usepackage{amsmath}		
\usepackage{cleveref}       
\usepackage{mathtools}
\usepackage[dvipsnames]{xcolor}
\usepackage{subcaption}

\usepackage{longtable}
\usepackage{multirow}
\usepackage{array}
\usepackage{nowidow}
\usepackage[ruled,lined,linesnumbered, noend]{algorithm2e}
\DontPrintSemicolon{}
\newcommand{\state}[1]{$#1$ \;}
\newcommand{\assign}[2]{\state{#1 \es{} \leftarrow{} \es{} #2}}
\newcommand{\es}{\textbf{ }}

\usepackage{enumitem}

\newcommand{\tabmargin}{0.25em}
\newcolumntype{P}[1]{>{\hspace{\tabmargin}}p{#1}<{\hspace{\tabmargin}}}
\newcolumntype{L}{>{\hspace{\tabmargin}}l<{\hspace{\tabmargin}}}
\newcolumntype{C}{>{\hspace{\tabmargin}}c<{\hspace{\tabmargin}}}
\newcolumntype{R}{>{\hspace{\tabmargin}}r<{\hspace{\tabmargin}}}

\title{\large\textbf{Compressing Hypergraphs using Suffix Sorting}} 

\author{Enno Adler,
Stefan Böttcher,
Rita Hartel\\[0.5em]
    {
    \small\begin{minipage}{\linewidth}
        \centering
            Paderborn University,\\
            Paderborn, Germany\\
            {\href{mailto:enno.adler@uni-paderborn.de}{\{enno.adler, stefan.boettcher, rita.hartel\}@uni-paderborn.de}}\\[0.5em]
    \end{minipage}
    }
}

\newcommand{\bigO}{\mathcal{O}}

\usepackage{color}

\begin{document}

\maketitle

\begin{abstract}
    Hypergraphs model complex, non-binary relationships such as co-authorships, social group memberships, and recommendation systems. Like traditional graphs, hypergraphs can grow large, posing challenges for storage, transmission, and query performance. We propose HyperCSA, a novel compression method for hypergraphs that maintains support for standard queries over the succinct representation. HyperCSA achieves compression ratios of $26\%$ to $79\%$ of the original file size on real-world hypergraphs—outperforming existing methods on all large hypergraphs in our experiments. HyperCSA also scales to larger datasets than existing approaches and evaluates neighbor queries $6$ to $40$ times faster on common real-world hypergraphs.
\end{abstract}


\section[Introduction]{Introduction}

Hypergraphs are powerful tools for modeling complex, non-binary relationships in data. Unlike traditional graphs, where edges connect only two nodes, hypergraphs allow edges to connect any number of nodes, enabling a more accurate representation of real-world systems. This modeling flexibility has made hypergraphs useful in a range of domains, including social networks~\cite{tan2014}, music~\cite{tan2011} and news~\cite{li2013} recommendation systems, academic authorship, and tagged data from online platforms~\cite{ghoshal2009}. Despite the widespread use of hypergraphs to store data, there is little research on succinct data structures for them. In this paper, we introduce the first self-index for hypergraphs called HyperCSA.  

Herein, our main contributions are as follows: 
\begin{itemize}[nolistsep]
    \item An algorithm and data structure, HyperCSA, to compress hypergraphs.
    \item Theoretical bounds on compression size, compression speed, and query performance for HyperCSA.
    \item A thorough experimental analysis of compression ratio, construction time, construction memory usage, and query performance of HyperCSA.
\end{itemize}

\section[Related Work]{Related Work} \label{section:related_work}

Like graphs, hypergraphs can be very large. For large hypergraph databases, there is the research branch of partitioning a hypergraph~\cite{schlag2022} and of parallel graph processing with tools like CHGL~\cite{jenkins2018} and Hygra~\cite{shun2020}. CHGL uses adjacency lists to represent hypergraphs: each node has an array of incident edges, and each edge has an array of incident nodes. Hygra~\cite{shun2020} uses the graph representation of Ligra~\cite{shun2020} on the bipartite representation of a hypergraph, which is also called star expansion, where each hyperedge is represented as a node connected by an edge to each node of the hyperedge. Besides the bipartite representation, each hypergraph can be translated into a traditional graph using the clique expansion, where a rank-k hyperedge is represented as a k-clique. This leads to a loss of information as overlapping cliques are indistinguishable or cease to exist. Both expansions have the downside that they increase the number of edges in the graph by a large factor in comparison to using hyperedges. Thereby, both expansions render the hypergraph less effective at compressing and need many edge lookups to check for an existing hyperedge between multiple nodes. Moreover, Huang et al.~\cite{huang2015} conclude in the context of distributed hypergraph processing, expanding hypergraphs into graphs causes major efficiency drawbacks.

Liu et al.~\cite{liu2024} argue likewise for graph processing acceleration. They propose a hypergraph compression method based on the incidence list: They use one array for the incident hyperedges, one array for the nodes, and four indexes, two for each of the arrays. Liu et al. achieve compression by overlapping the common nodes (or common hyperedges) of successive hyperedges (or successive nodes). The indexes contain the start and end of the possibly overlapping ranges of adjacent nodes or hyperedges. As neither nodes nor hyperedges have an order, they propose a reordering strategy for better compression and less cache misses on access.

Our approach, HyperCSA, is inspired by RDFCSA by Brisaboa et al.~\cite{brisaboa2015}, a self-index for the triples representing an RDF graph. We generalize their idea by lifting the restriction to triples and permit tuples of arbitrary and unequal lengths.


\section[Preliminaries]{Preliminaries}

\begin{figure*}[!hbp]
    \centering
    \includegraphics[scale=0.8]{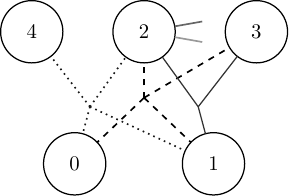}
    \caption{The example hypergraph $H$ with nodes $V_H = \{0, 1, 2, 3, 4\}$ and hyperedges $E_H = \left\{\{0, 1, 2, 3\},\es\{1, 2, 3\},\es\{2\},\es\{0, 1, 2, 4\},\es\{2\}\right\}$.}\label{figure:example_hypergraph}
\end{figure*}

A \textit{hypergraph} is a pair $G = (V_G, E_G)$ with nodes $V_G = \{0, 1, \ldots, p\} \subset \mathbb{N}_0$ and edges $E_G = \{e_0, \dots, e_{m-1}\}$. For a hyperedge $e_i = \{v_0, \ldots, v_{n-1}\} \subseteq V_G$, $e_i \not= \emptyset$ and $rank(e_i) = |e_i| = n$. We call all hyperedges in $E_G$ edges regardless of their rank. 
HyperCSA can also handle hypergraphs that contain the same edge multiple times. In \autoref{figure:example_hypergraph}, we visualize our ongoing example.

We define $degree(v) = |\{e \in E_G : v \in e \}|$, $R = \max\{rank(e): e\in E_G\}$, $\mathcal{V} = \max\{degree(v):v\in V_G\}$, and $M_G = \sum_{e \in E_G} rank(e) = \sum_{v \in V} degree(v)$. For the example hypergraph $H$, we have $degree(2) = 5$, $R = 4$, $\mathcal{V} = 5$, and $M_H = 13$.

A string $T$ of length $|T|$ over an alphabet $\Sigma$ is a sequence of characters $T[i] \in \Sigma$ for $0 \leq i < |T|$, so $T = T[0] \cdot T[1] \cdots T[|T|-1]$. 
We write $T[i, j]$ for the substring $T[i] \cdot T[i+1] \cdots T[j]$ of $T$, $T[i,j] = \epsilon$ if $i > j$, and $T[i..] = T[i, |T|-1]$ for the suffix starting at position $i$. We call a position $i$ an S-type, if $T[i] < T[i+1]$, and L-type otherwise, also if $i=|T|-1$.

We define $rank_T(x, c) = |\{i \leq x : T[i] = c\}|$ as the number of times $c$ occurs in the first $x+1$ characters of $T$ (including the current position). 
Next, let $select_T(x, c) = \min\{i < |T| : rank_T(i, c) = x\}$ be the position of the $x$-th $c$ in $T$. 

The suffix array $SA$~\cite{manber90} of a string $T$ is a permutation of $\{0, \ldots, |T|-1\}$ such that the $i$-th smallest suffix of $T$ is $T[SA[i]..]$ with $i \in \{0, \ldots, |T|-1\}$. As permutations are invertible, we denote the inverse of the suffix array by $SA^{-1}$. We use the term \textit{T-position} for a position in $T$ and \textit{SA-position} for a position in $SA$.

Sadakane's~\cite{sadakane2003} compressed suffix array $CSA$ of a string $T$ consists of a bitvector $D$ of length $|T|$ and a permutation $\Psi$ of $\{0, \ldots, |T|-1\}$ (and other here omitted data structures). 

$D$ is a succinct representation for $T$ using the suffix array $SA$: $D[i] = 1$ marks the first suffix in the sorted suffixes of $T$ that start with a new character, so $D[i] = 1$ if and only if $T[SA[i]] > T[SA[i-1]]$ or $i = 0$. We assume that all symbols of $\Sigma$ have at least one occurrence in $T$. By this assumption, we get $T[SA[i]] = rank_D(i, 1)-1$.

$\Psi[i]$ stores the SA-position $j$ that points to the next T-position $SA[i]+1$ after the T-position $SA[i]$ at SA-position $i$. In one term, $\Psi[i] = SA^{-1}[(SA[i] + 1) \mod |T|]$. $\Psi$ is a cyclic permutation and by computing the ranks on $D$ for the SA-positions $i$ in order of the cycle of $\Psi$, we retrieve a rotation of $T$. 

\section[HyperCSA]{HyperCSA}\label{section:hypercsa}


\subsection{Construction}

The outline of HyperCSA's construction is as follows: First, because we use a CSA as the main component to compress the hypergraph $G$, we encode $G$ as a string $T$. Second, we build the CSA for $T$. Third, we adjust $\Psi$ such that each edge $e \in E_G$ has its own cycle in $\Psi$. This way, repeated applications of $\Psi$ stay within $e$ and we get $e$ by traversing through its cycle.

In the first step, we construct a string $T$ of length $M_G$ over the alphabet $V_G$. We begin by sorting the nodes within each edge in ascending order; now, because each edge $\{v_0, \dots , v_{n-1}\}$ can contain each node at most once, we get $v_0 < v_1 < \dots < v_{n-1}$. 
Second, we sort the edges in $E_G$ in descending lexicographic order. If one edge $e_p$ is a complete prefix of another edge $e_q$, we set $e_p < e_q$. For two edges $(v_0, \dots, v_{n-1}) > (w_0, \dots, w_{l-1})$, we especially have $v_0 \geq w_0$ and thus $v_{n-1} \geq w_0$. 
Then, we copy each node from each edge to $T$ according to the sort order. 
In the example of \autoref{figure:example_hypergraph}, the resulting string is $T = 2 | 2 | 1 2 3 | 0 1 2 4 | 0 1 2 3$. We can reconstruct $E_G$ from $T$ by partitioning $T$ after each L-type position $i$. This is because $T[i]$ is S-type if and only if node $T[i+1]$ belongs to the same edge as $T[i]$. 


Second, we build the CSA for $T$. 
The bitvector $D$ is the unary encoding of $degree(v)$ for the nodes $v \in V_G$. 
Thus, we construct $D$ by counting the frequency of nodes in $T$ in an array $freq$ of length $|V_G|$. Then, we set $D\left[\sum_{i = 0}^{j-1} freq[i]\right]=1$ for every $j = 0, \dots, |V_G|-1$ and $D[l]$ to $0$ at all other positions $l$. Additionally, we append a $1$ at the end, so we build the bitvector $D$ of length $M_G+1$ instead of length $M_G$. 
By adding the $1$, we can always use $select$ to find the interval of $D$ associated to a node $v \in V_G \subset \mathbb{N}_0$: The interval $S(v) = [select_D(v+1, 1), select_D(v+2, 1))$ contains the SA-positions $i$ where $T[SA[i]] = v$. We call $select$ with $v+1$ instead of $v$ because $D[0] = 1$ and $rank_D(0, 1) = 1$, but we associate the position $0$ in $D$ with node $0$. Without adding the $1$ at the end of $D$, the expression $select(v+2, 1)$ for $v=|V_G|-1$ would become undefined. 

\begin{figure*}
    \centering
    \includegraphics[width=0.8\linewidth]{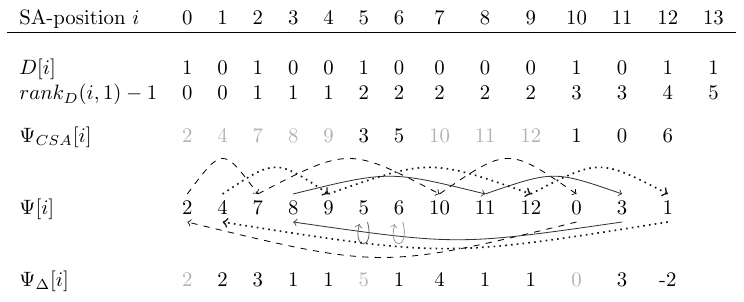}
    \caption{HyperCSA data structure for the hypergraph $H$. $rank_D(i, 1)-1$ is the node at index $i$. The black numbers in $\Psi_{CSA}$ are the numbers adjusted by \autoref{algo:adjust_psi}. The style of every edge in $\Psi$ is the same as in \autoref{figure:example_hypergraph}. The SA-positions with $\Psi[i] \leq i$ are below the numbers, all others are above. The black numbers in $\Psi_\Delta$ are $\Psi[i] - \Psi[i-1]$. We use a distance of $t = 5$ in $\Psi_\Delta$.}
    \label{figure:psi_jumps}
\end{figure*}

In our example, we obtain $D = 1 0 1 0 0 1 0 0 0 0 1 0 1 1$, because two edges are incident to node $0$, three edges to node $1$, five edges to node $2$, two edges to node $3$, and only one edge to node $4$. In \autoref{figure:psi_jumps}, we show $D$ as well as the corresponding node $rank_D(i, 1)-1$ for every SA-position of $D$.

In the third step, we adjust the cyclic permutation $\Psi_{CSA}$. We refer with $\Psi_{CSA}$ to the permutation constructed by the CSA and with $\Psi$ to the permutation that we compute. Let $i$ be the SA-position of the T-position $j = SA[i]$. Proceeding from $T[j]$ to $T[j+1]$, which corresponds to applying $\Psi_{CSA}$ on $i$, we can have the two cases: 

\begin{itemize}
    \item $j$ is an S-Type, so $T[j+1]$ is a node of the same edge as $T[j]$. Then, $\Psi_{CSA}[i] > i$.
    \item $j$ is an L-Type, so $T[j+1]$ is the first node of a new edge. Then, $\Psi_{CSA}[i] < i$. 
\end{itemize}

A position $i$ with $\Psi_{CSA}[i] = i$ forms a cycle of length $1$, which contradicts that $\Psi_{CSA}$ consists only of a single cycle of length $M_G$. Thereby, $\Psi_{CSA}[i] \neq i$ for all SA-positions $i$. Next, we want to adjust $\Psi_{CSA}$ in such a way that $\Psi$ contains one cycle for each edge in $E_G$. Since $\Psi_{CSA}$ shows how to reassemble $T$, we only need to adjust the T-positions $k$ where $k$ is an L-type.

Let $i$ be an L-type SA-position 
and let $k, k+1, \dots, k+n-2, k+n-1$ be the T-positions of the edge $e$ of rank $n$, where $SA[i] = k+n-1$. For $\Psi[i]$, we need the SA-position $j$ such that $SA[j] = k$. Then, we set $\Psi[i] = j$ and thereby form the cycle for the edge $\{T[k], T[k+1], \dots, T[k+n-1]\}$. For all other S-type SA-positions, 
we set $\Psi[i] = \Psi_{CSA}[i]$.

To compute $\Psi$, we could iterate over $\Psi_{CSA}[i]$ by SA-positions $i$ to find the SA-positions where $\Psi_{CSA}[i] < i$. However, we would not know which SA-position $j$ to assign to $\Psi[i]$. Instead, we iterate by T-positions $i$ and follow the cycle of $\Psi_{CSA}$: The last SA-position $\Psi_{CSA}[k]$ with $\Psi_{CSA}[k] < k$ (L-type) points towards the SA-position $\Psi_{CSA}[k] = j$. So, if $i$ is the next L-type SA-position with $\Psi_{CSA}[i] < i$ after $\Psi_{CSA}[k] < k$, we set $\Psi[i]$ to be $j$.

If we iterate by T-positions $i$, we use the sequence of SA-positions $(\Psi_{CSA})^{i}[0]$, because applying $\Psi_{CSA}$ corresponds to increasing the position $i$ in $T$ by one. 
We start at position $0$ in the suffix array because $\Psi_{CSA}$ is a permutation and forms a cycle of length $M_G$, the SA-position $i$ with $\Psi_{CSA}[i] = 0$ must be greater than $0$, so $\Psi[i] = 0 < i$ and $i$ is an L-type. This means that $0$ is the SA-position of a first node of an edge. 

The steps to adjust $\Psi_{CSA}$ to $\Psi$ together yield \autoref{algo:adjust_psi}. We visualized these cycles for our example in \autoref{figure:psi_jumps}. 
\begin{algorithm}
    \caption{\(adjust\Psi\)}\label{algo:adjust_psi}
    \KwIn{$\Psi$}
    \assign{current\_position}{\Psi[0]}
    \assign{next}{\Psi[current\_position]}
    \assign{last\_first\_node\_position}{0}
    \While{$current\_position \neq 0$}{
        \If{$next < current\_position$}{
            \assign{\Psi[current\_position]}{last\_first\_node\_position}
            \assign{last\_first\_node\_position}{next}
        }
        \assign{current\_position}{next}
        \assign{next}{\Psi[next]}
    }
\end{algorithm}

\subsection{Decompression}

Every SA-position $i < M_G$ belongs to exactly one edge $e$, so given an SA-position $i$, we can extract the corresponding edge $e$ as follows: We calculate every SA-position $j$ with $j = \Psi^k[i]$ for $k = 0, 1, \dots$ until $\Psi^k[i] = i$, which happens the first time for $k = rank(e)$. We then use $n = rank_D(j, 1)-1$ for every such SA-position $j$ to determine the node $n$. In one expression, $e = \{rank_D(\Psi^k[i], 1)-1 : k = 0, 1, \dots\}$.

For the proper decompression, we now need one position per edge. By construction, every edge has exactly one L-type position $i$. Thereby, it is sufficient to scan $\Psi$ for all positions $i$ with $\Psi[i] \leq i$ and decompress the edge at $i$. Thus, \[E_G = \left\{ \{rank_D(\Psi^k[i], 1)-1 : k = 0, 1, \dots\} : \Psi[i] \leq i \right\}.\]

\subsection{Theoretical Analysis of HyperCSA}

$D$ has $M_G+1$ bits and we store $D$ as a plain bitvector. We add data structures requiring $o(M_G)$ space that support rank~\cite{vigna2008} and select~\cite{clark1997} queries in $\bigO(1)$ time.

$\Psi$ can be encoded significantly smaller than $M_G$ integers, because $\Psi$ consists of at most $|V_G|$ strictly increasing subsequences. Thereby, we use the differences $\Psi_\Delta[i] \coloneqq \Psi[i] - \Psi[i-1]$. $\Psi_\Delta$ needs space proportional to the zero-order empirical entropy $H_0(T)$ of $T$~\cite{ferragina2010permuterm}: $\Psi_\Delta$ can be represented in $M_G H_0(T) + \bigO(M_G \log(H_0(T))) \leq M_G \log(|V_G|) + o(M_G \log(\log(|V_G|)))$ bits. Moreover, using $\delta$-codes and run-length encoding for runs of $1$ in $\Psi_\Delta$, the size of $\Psi_\Delta$ gets closer to higher order entropy~\cite{brisaboa2015,navarro2007}. 
For fast practical access, $\Psi_\Delta$ stores the absolute value of $\Psi$ at every position that is a multiple of $t = 128$. This ensures a constant access time of $\bigO(t)$. In \autoref{figure:psi_jumps}, we show $\Psi_\Delta$ for our example with $t = 5$.



The total construction time of HyperCSA is $\bigO(|E_G| R (\log(R) + \log(|E_G|)))$. First, we do $|E_G|$ sorting steps of nodes in edges, which takes $\bigO(|E_G| R \log(R))$ time. Second, sorting the edges takes $\bigO(|E_G| R \log(|E_G|))$ time, because a single comparison of two edges can take up to $R$ steps. Constructing $D$ takes $\bigO(M_G)$ time for the frequency scan of $T$ and $\bigO(M_G)$ time for assigning the unary code to $D$. Constructing the CSA, which includes $\Psi_{CSA}$, takes $\bigO(M_G)$ time. Finally, \autoref{algo:adjust_psi} traverses the cycle of $\Psi_{CSA}$ once, which also takes $\bigO(M_G)$ steps. 


\subsection{Queries}


\paragraph{\boldmath$degree(v)$:} The $degree$ query returns the number of edges incident to a node $v\in V_G$. As we can compute $select$ in $\bigO(1)$ time~\cite{clark1997} and $degree(v) = select_D(v+2, 1) - select_D(v+1, 1)$, $degree(v)$ only takes $\bigO(1)$ time.

\paragraph{\boldmath$contains(v_0, \dots, v_{k-1})$:} The $contains$ query returns all edges that contain all the nodes $v_0$ to $v_{k-1}$, so $contains(v_0, \dots, v_{k-1}) = \{e \in E_G : \{v_0, \dots, v_{k-1}\} \subseteq e\}$. We assume that the nodes of the query are in ascending order. Let $v_{m}$ be the node of the query with the smallest $S(v_i)$ for all $i=0, \dots, k-1$. Every position $i \in S(v_{m})$ belongs to one (different) edge that contains the node $v_{m}$.

We test each edge at a position $i \in S(v_{m})$ individually if the set $\{\Psi^k[i] : k = 0, 1, \dots \}$ contains the position of the other nodes of the query. We sorted the query because we traverse $\Psi^k[i]$ and the nodes will appear in the same order in the edge as in the sorted query. However, the edge is allowed to contain more nodes than the query. When the last node of the query $v_{m -1 \mod k}$ is found in an edge, we can output the edge as a valid result. Note that we partially decompressed the edge for checking the nodes, so we only need to decompress the remaining nodes of the edge.

We have multiple conditions to abort and discard an edge:
\begin{enumerate}[nolistsep]
    \item If $\Psi^l[i] = i$ for $l\in\mathbb{N}$ before we found all nodes of the query.
    \item If the current node $rank_D(\Psi^l[i], 1)-1$ is higher than the next node of the query, except if the next node of the query is the lowest node of the query.
    \item If we are at the L-type position $j$ with $\Psi[j] < j$ for the edge and the next node of the query is not the lowest node.
    \item If we look for the lowest node of the query, we are at position $j = \Psi^l[i]$ with $\Psi[j] < j$, and the node $rank_D(\Psi^{l+1}[i], 1)-1$ is greater than the node of the query.
\end{enumerate}
All these discard conditions originate from the order of the nodes in an edge and the cyclic structure of $\Psi$.

Overall, for the $contains$ query, we search the minimum over $k$ values. Then, for each edge of that interval, we perform at most $R$ decompression steps. Each test of the next query node costs only $\bigO(1)$. Thus, we have a runtime of $\bigO(k + \min(degree(v_0, \dots, v_{k-1})) \cdot R)$.

\paragraph{\boldmath$exists(v_0, \dots, v_n)$:} The $exists$ query answers how many edges $\{v_0, \dots, v_n\}$ exist in $E_G$, so $exists(v_0, \dots, v_n) = |\{\{v_0, \dots, v_n\} \in E_G\}|$. In the case of $E_G$ being a multiset, the answer might be larger than one. 

Again, we assume $v_0 < \dots < v_n$. First, we set $\mathcal{I}_0 = S(v_0)$. Second, we iterate over the nodes $v_2$ to $v_n$. At the iteration of node $v_i$, we determine $\mathcal{I}_i$ using $\mathcal{I}_{i-1}$, where each $\mathcal{I}_i$ is a subinterval of $S(v_i)$ as follows: We binary search the subinterval $\mathcal{S}$ of $\mathcal{I}_{i-1} \subseteq S(v_{i-1})$ with the property that for $i \in \mathcal{S}$, $\Psi[i] \in S(v_{i})$. Then, we set $\mathcal{I}_i = \Psi[\mathcal{S}] = [\Psi[\min(\mathcal{S})], \Psi[\max(\mathcal{S})]]$. Third, we have $\mathcal{I}_n$ for the highest node $v_n$ now. We check for closed cycles like in the previous steps: We search the interval $\mathcal{S}$ of $\mathcal{I}_{n}$ with $\Psi[i] \in S(v_{0})$. Then, we output the length of the interval $[\Psi[\min(\mathcal{S})], \Psi[\max(\mathcal{S})]]$ as the result to the $exists$ query. 

For the $exists$ query with $v_0, \dots, v_n$, we do $2(n+1)$ binary searches on the intervals or subintervals of $S(v_i)$. Because we answer $select$ queries on $D$ in $\bigO(1)$, we can get $S(v)$ for $v\in V$ in $\bigO(1)$. The length of $S(v)$ is $degree(v)$. Thereby, the binary search on $\Psi$ on the interval $S(v)$ takes only $\bigO(\log(degree(v)))$. Thus, the runtime of an exists query is $\bigO(\sum_{i=0}^{n}\log(degree(v_i)))$. 


\section{Experimental Results}

We compare the hypergraph representations in \autoref{tab:algorthims} on the hypergraphs listed in \autoref{tab:data_sets}. We investigate midsize hypergraphs (MaAn, WaTr, and TrCl), and large hypergraphs (CoFr, StAn, AmRe, and CoOr).

\begin{table}
    \centering
    \begin{tabular}{lll}
        approach & paper & implementation\\
        \hline

        \href{https://github.com/adlerenno/hypercsa}{HyperCSA} & this paper & \url{https://github.com/adlerenno/hypercsa} \\

        \href{https://github.com/jshun/ligra}{Hygra} & \cite{shun2020} & \url{https://github.com/jshun/ligra} \\

        \href{https://github.com/muranhuli/Reordering-and-Compression-for-Hypergraph-Processing}{Array-vanilla} & \cite{liu2024} & \multirow{3}{*}{\parbox{5cm}{\url{https://github.com/muranhuli/Reordering-and-Compression-for-Hypergraph-Processing}}} \\

        \href{https://github.com/muranhuli/Reordering-and-Compression-for-Hypergraph-Processing}{Array-V} & \cite{liu2024} \\

        \href{https://github.com/muranhuli/Reordering-and-Compression-for-Hypergraph-Processing}{Array-E} & \cite{liu2024} \\

        \href{https://github.com/muranhuli/Reordering-and-Compression-for-Hypergraph-Processing}{Array-V\&E} & \cite{liu2024} \\

        \href{https://github.com/muranhuli/Reordering-and-Compression-for-Hypergraph-Processing}{incidence list} & \cite{liu2024} \\

        \href{https://github.com/adlerenno/hypercsa-test/blob/main/scripts/incidence_matrix.py}{incidence matrix} & – & \url{https://github.com/adlerenno/hypercsa-test}\\
    \end{tabular}
    \caption{Used hypergraph representations.}\label{tab:algorthims}
\end{table}

To evaluate practical applicability, we measure\footnote{The test is available at \url{https://github.com/adlerenno/hypercsa-test}.} construction time, compression ratio, RAM usage, and query performance. All tests were performed on a Debian 5.10.209-2 machine with 512GB RAM and 128 Cores Intel(R) Xeon(R) Platinum 8462Y+ @ 2.80GHz. We implemented HyperCSA using the Succinct Data Structure Library~\cite{gog2014}. We were unable to evaluate CHGL due to outdated package requirements, but CHGL uses an incidence list implementation comparable to the included implementation of Liu et al.~\cite{liu2024}. 

\begin{table}
    \centering
    \addtolength{\tabcolsep}{-0.2em}
    \begin{tabular}{p{4.8cm}>{\hspace{0.8em}}rrrrr}
        hypergraph $G$ & $|V_G|$ & $|E_G|$ & $M_G$ & $R$ & $\mathcal{V}$ \\ \hline





        \makebox[0pt][l]{\href{https://www.cs.cornell.edu/~arb/data/mathoverflow-answers/}{mathoverflow-answers} (MaAn)} & 73,851 & 5,446 & 131,714 & 1,784 & 173 \\

        \href{https://www.cs.cornell.edu/~arb/data/walmart-trips/}{walmart-trips} (WaTr) & 88,860 & 69,906 & 460,630 & 25 & 5,733 \\ 

        \href{https://www.cs.cornell.edu/~arb/data/trivago-clicks/}{trivago-clicks} (TrCl) & 172,738 & 233,202 & 726,861 & 85 & 339 \\

        \href{https://snap.stanford.edu/data/com-Friendster.html}{com-friendster} (CoFr) & 7,944,949 & 1,620,991 & 23,479,217 & 9,299 & 1,700 \\

        \makebox[0pt][l]{\href{https://www.cs.cornell.edu/~arb/data/stackoverflow-answers/}{stackoverflow-answers}$\,$(StAn)} & 15,211,989 & 1,103,243 & 26,109,177 & 61,315 & 356 \\ 

        \href{https://www.cs.cornell.edu/~arb/data/amazon-reviews/}{amazon-reviews} (AmRe) & 2,268,231 & 4,285,363 & 73,141,425 & 9,350 & 28,973 \\

        \href{https://snap.stanford.edu/data/com-Orkut.html}{com-orkut} (CoOr) & 2,322,299 & 15,301,901 & 107,080,530 & 9,120 & 2,958 \\ 
    \end{tabular}
    \caption{Used datasets from \href{https://snap.stanford.edu/data/index.html}{SNAP}~\cite{yang2012} and \href{https://www.cs.cornell.edu/~arb/data/}{ARB} sorted by the value of $M_G$.
    }\label{tab:data_sets}
\end{table}

\begin{figure}
    \centering
    \includegraphics[width=0.9\textwidth]
    {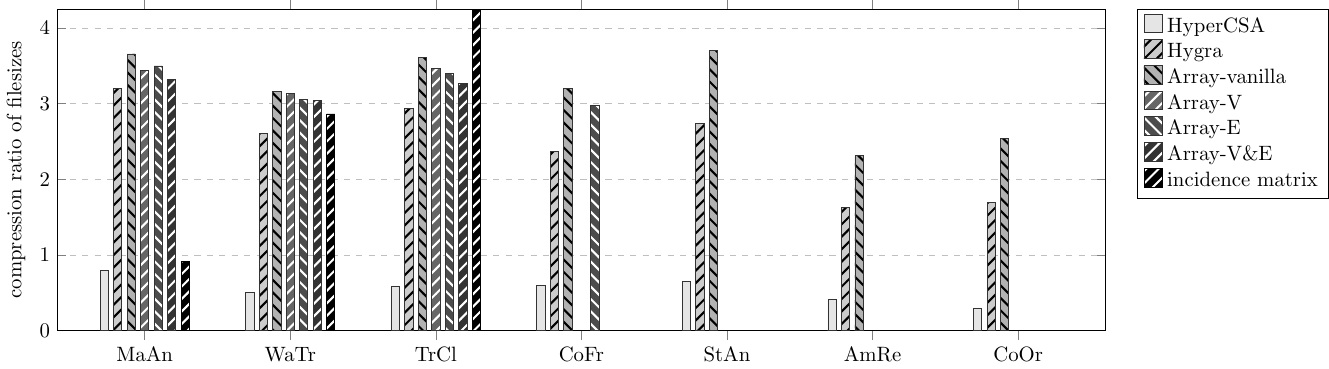}
    \caption{Hypergraph compression sizes in relation to the original file size. }\label{figure:compression_ratio}
\end{figure}

First, we evaluate compression ratios in \autoref{figure:compression_ratio}. Compression ratios above one indicate output sizes that are larger than the plain list input file. For large hypergraphs, missing data points of Array-V, Array-E and Array-V\&E indicate abortions due to execution times of more than 5 days each. 
On all hypergraphs, HyperCSA achieves the best compression. HyperCSA is at least four times smaller in compression ratio than the approach of Liu et al.~\cite{liu2024} and also at least four times smaller than Hygra. The incidence matrix is compact for the MaAn hypergraph but infeasible on large hypergraphs like CoOr, where it exceeds $35$Tb of memory. Overall, HyperCSA provides consistently superior compression. 

Second, we investigate resource costs of the compression. 
Array-vanilla has the fastest construction with always less than $9$ minutes. Nevertheless, HyperCSA stays below $4$ minutes for most hypergraphs, only taking $36$ minutes on AmRe and $51$ minutes on CoOr—mainly for the construction of $\Psi_{CSA}$. 
Except for the incidence matrix, all representations use less than 8GB of RAM. 

Third, we evaluate query times. For every dataset and query type, we generated 1000 queries. The presented values are the average runtime. 

\begin{figure}
    \centering
    \begin{subfigure}[t]{0.59\textwidth}
        \setcounter{subfigure}{0}
        \centering
        \includegraphics[width=\textwidth]{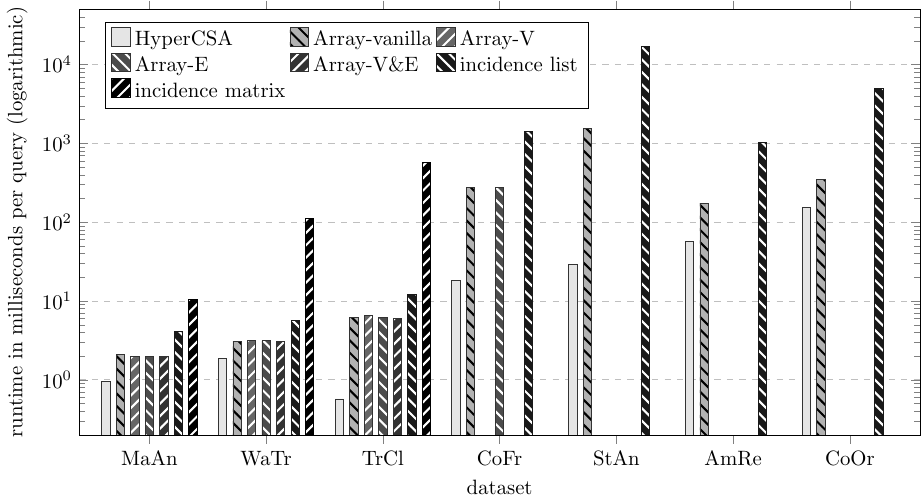}
        \vfill
        \begin{minipage}{0.82\linewidth}
            \caption{Query times of $contains(v_0)$ query, which returns all neighbors of node $v_0$.}\label{figure:query_times}
        \end{minipage}
    \end{subfigure}
    \hfill
    \begin{subfigure}[t]{0.39\textwidth}
        \centering
        \includegraphics[width=\textwidth]{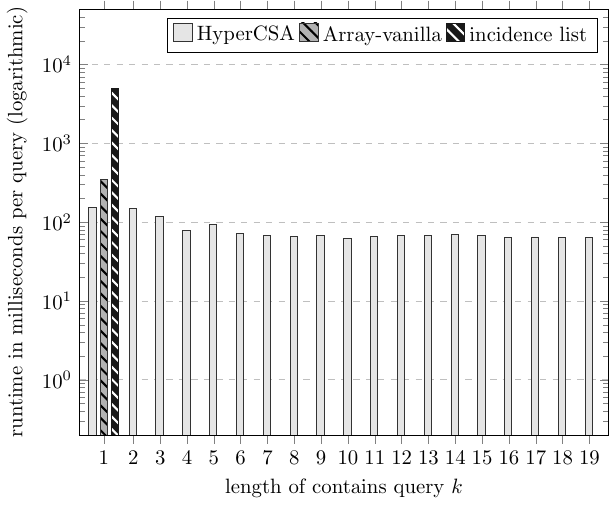}
        \begin{minipage}{0.805\linewidth}
            \caption{Query times of $contains(v_0, \dots v_{k-1})$ query on the largest dataset CoOr.}\label{figure:higher_contain_queries}
        \end{minipage}
    \end{subfigure}
    \caption{Query time of $contains$ query. Missing datapoints indicate that the method failed to complete in decent time.}
\end{figure}

For $contains(v_0)$ queries with a single node, which correspond to neighbor queries on hypergraphs, the results in \autoref{figure:query_times} show a performance gap, with HyperCSA outperforming the other approaches by being $6$ to $40$ times faster.


For $contains(v_0, \dots, v_{k-1})$ queries, \autoref{figure:higher_contain_queries} presents the performance on the largest dataset, CoOr. Only HyperCSA is capable of performing these queries. HyperCSA answers these queries in about $60$ milliseconds. As the number of nodes $k$ increases, it becomes more likely that exactly one edge matches the $contains$ query, decreasing the time to decompress the edges. 

Only HyperCSA and the incidence matrix can perform $exists$ queries. HyperCSA uses only $0.2$ milliseconds to $0.66$ milliseconds per query on the datasets MaAn, WaTr, and TrCl. Thereby, HyperCSA outperforms the incidence matrix by a factor of $250$ to $850$. 

\section{Conclusion}

We introduced HyperCSA, a succinct and scalable data structure for compressing and querying hypergraphs. HyperCSA employs multiple sorting steps and compressed suffix arrays and uses this sorting beneficially for querying the hypergraph. Our experiments demonstrate that HyperCSA consistently outperforms existing approaches, achieving better compression ratios and faster query times of up to an order of magnitude on large-scale, real-world hypergraphs.

\Section{References}
\bibliographystyle{IEEEtran}
\bibliography{biblio}

\end{document}